\documentclass[11pt,a4paper,english,nofootinbib,,superscriptaddress]{revtex4}
\usepackage{lmodern}

\usepackage[T1]{fontenc}
\usepackage[latin9]{inputenc}
\usepackage{babel}
\usepackage{amsmath}
\usepackage{graphicx}
\usepackage{amssymb}
\usepackage{esint}
\usepackage[unicode=true, pdfusetitle,
 bookmarks=true,bookmarksnumbered=false,bookmarksopen=false,
 breaklinks=false,pdfborder={0 0 1},backref=false,colorlinks=false]
 {hyperref}
\usepackage{amsmath}
\usepackage{amssymb}
\def\b{\begin{equation}}
 \def\e{\end{equation}}

\makeatletter
\usepackage{latexsym}\usepackage{bm}

\makeatother

\begin{document}

\title{An Asymptotic Method for Selection of Inflationary Modes}

\author{E. Yusofi}
\email{e.yusofi@iauamol.ac.ir}
\affiliation{Department of Physics, Ayatollah Amoli Branch, Islamic Azad University, Amol, Mazandaran, Iran}
\author{M. Mohsenzadeh}
\email{mohsenzadeh@qom-iau.ac.ir}
\affiliation{Department of Physics, Qom Branch, Islamic Azad University, Qom, Iran}

\date{\today}

\begin{abstract}

\noindent \hspace{0.35cm} We present some features of early
universe cosmology in terms of Hankel functions index ($\nu$).
Actually, the recent data from observational cosmology indicate
that our universe was nearly de Sitter space-time in the early
times which results in an approximate scale-invariant spectrum.
This imposes some constrains on index $\nu$ \cite{man1}. These
constrains stimulate us to use general solution of
 inflaton field equation for $\nu\neq{\frac{3}{2}}$.  To obtain the general solution for the inflationary background, we use
 asymptotic expansion of Hankel functions up to non-linear order of $\frac{1}{k\eta}$. We consider the non-linear modes as
 the fundamental modes for early universe during the inflation. In this paper, we obtain the general form
 of the inflationary modes, scale factor expansion, equation of state and some non-linear corrections of power spectrum in terms of index $\nu$.
 These results are general and in the quasi- de Sitter and de Sitter limit confirm the conventional results.\\

\textbf{Pacs}: 04.62.+v, 04.25.-g, 98.80.Cq\\

\end{abstract}

\maketitle

\section{Introduction}

The important formulation of two linearly independent solutions of
Bessel's equation are Hankel functions $H_{\nu}^{(1)}(x) $ and $
H_{\nu}^{(2)}(x)$ , which are defined by,
$$ H_{\nu}^{(1)}(x)=B_{\nu}(x)+iN_{\nu}(x)$$
\begin{equation}
H_{\nu}^{(2)}(x)=B_{\nu}(x)-iN_{\nu}(x),
\end{equation}
where $i=\sqrt{-1}$. These linear combinations are also known as
\emph{Bessel functions of the third kind}. In this equation the
$B_{\nu}(x)$ and $N_{\nu}(x)$ are the Bessel functions of the
first and second kind, respectively \cite{tra11, tra12}. The Hankel
functions are used to express outward and inward propagating
cylindrical or spherical wave solutions of the wave equation,
depending on the positive or negative sign of the frequency. Also,
in the cosmology they are used to express the general solution of
the inflaton field equation \cite{inf2, inf3, inf41, inf42, per5, per6}.\\
The recent CMB results from Planck satellite and data from the
Wilkinson Microwave Anisotropy Probe (WMAP), impose an important
constraint on the value of scalar spectral index approximately to
be  $n_s = {0.9603} \pm {0.0073}$ at 95\% CL \cite{obs71, obs72, obs8}. In
\cite{man1}, we considered this constraint to show that the index
of Hankel function, lies in the range of $1.51\leq \nu \leq 1.53$.
The usual Bunch-Davies (BD) mode \cite{Bun10} is used just for the
pure de Sitter(dS) space-time i.e., it is obtained by exactly
setting $\nu={\frac{3}{2}}$, thus the above range of index $\nu$
 motivates us to consider non-de Sitter (ND) modes with  $\nu\neq{\frac{3}{2}}$. To obtain the ND modes, we have used approximate method
 in \cite{man11, man12}, but in \cite{man1} and present work, we exploit the asymptotic expansion of Hankel functions up to the higher order
 of $\frac{1}{x}=\frac{1}{k\eta}$. We nominate the non-de Sitter modes as the fundamental modes for early universe and obtain the cosmic expansion, equation of state and power spectrum in terms of index $\nu$. \\
The rest of this paper proceed as follows: In Sec. 2, we first introduce equation of motion for inflaton field and its
general solution. Then by considering the asymptotic expansion of Hankel functions up to the non-linear order of $\frac{1}{k\eta}$,
 we obtain ND modes in terms of index $\nu$. In Sec. 3, we obtain some cosmological issues such as: the cosmic expansion, equation of
 state and non-linear corrections of power spectrum in terms of index $\nu$. Cosmological interpretations for some special values of
 index $\nu$ and conclusions will be discussed in the final section.

\section{General Solution of Mukhanov Equation}

The following metric is usually used to describe the universe with
isotropic expansion \cite{inf2}
\begin{equation}
ds^{2}=dt^{2}-a(t)^2{d\textbf{x}}^{2}=a(\eta)^2({d\eta}^2-{d\textbf{x}}^2),
\end{equation}
where the scale factor is defined by $a(\eta)$ is the function of
the conformal time $\eta$. There are some models of inflation, but
the single-field inflation in which a minimally coupled scalar
field (inflaton) in dS background, is well motivated in the
literatures. The action for single-field models leads to the
equation of motion for the mode functions $u_{k}$ (or Mukhanov
equation) \cite{per5, non29, non32},
\begin{equation} \label{muk7}  u''_{k}+(k^{2}-\frac{ z''}{z})u_{k}=0. \end{equation}
The general solutions of mode equation (\ref{muk7}) can be written
in terms of Hankel functions as \cite{inf3, inf41, inf42, per5, per6}:
\begin{equation} \label{Han22} u_{k}=\frac{\sqrt{\pi
\eta}}{2}\Big(A_{k}H_{\nu}^{(1)}(|k\eta|)+B_{k}H_{\nu}^{(2)}(|k\eta|)\Big),
\end{equation} where $ H_{\nu}^{(1)} $ and $ H_{\nu}^{(2)}$ are
the first and second kind of Hankel functions, respectively
\cite{inf3, per6}. But since the function  $z$ in mode equation
(\ref{muk7}) is a time-dependent parameter and depends on the
dynamics of the background space-time, thus finding the exact
solutions of the equation (\ref{muk7}) is difficult so that the
numerical or the approximation methods are usually used.

\subsection{Asymptotic Expansion of Hankel Functions as the Early Time Solution}

 We have used asymptotic expansions of Hankel function up to the higher
  order of $\frac{1}{|k\eta|}$ \cite{man1}, for the far past time $|k\eta|\gg1$ in the very early universe as follows \cite{per5, tra11, tra12, man51},

 $$ H_{\nu}^{(1, 2)}(|k\eta|)\rightarrow$$
 $$\sqrt{\frac{2}{\pi{|k\eta|}}}\big[1\pm{i}\frac{4\nu^2-1}{8|k\eta|}-\frac{(4\nu^2-1)(4\nu^2-9)}{2!(8|k\eta|)^2}\pm...\big]$$
  \begin{equation}
 \label{asm23}\times{exp}[\pm{i}(|k\eta|-(\nu+\frac{1}{2}))\frac{\pi}{2}].
\end{equation}
Note that, this asymptotic expansion, only for $\nu=\frac{3}{2}$
reduces to the pure dS mode which results in the first (linear)
order in terms of $\frac{1}{|k\eta|}$ and for another values of
index $\nu$, the modes contain non-linear order in terms of
$\frac{1}{|k\eta|}$. As we know the recent observational results
indicate that our universe starts in an approximate dS or quasi-dS
space-time \cite{obs71, obs72, obs8} with varying Hubble parameter, so one
can use $\nu\neq \frac{3}{2}$ and non-dS modes.

\section{Cosmological Applications}
\subsection{Cosmic Expansion and Equations of State in Terms of Index $\nu$}

For the dynamical inflationary background, in equation
(\ref{muk7}) one should set $\frac{{z''}}{z}\neq0 $  and it is a
time-dependent value in terms of the conformal time $\eta$
\cite{asl55},
\begin{equation} \label{zed24}
\frac{{z''}}{z}=\frac{4\nu^2-{1}}{4\eta^2}.
 \end{equation}
In addition to the variable $\eta $, the value of $
\frac{{z''}}{z}$ depends on Hankel function index $\nu$. The
general form of equations (\ref{zed24}) is \cite{tra11, tra12}, \b
\label{Spa35} {\eta}^2{z''}-Cz=0, \e the general solution of the
above equation in terms of conformal time $\eta$ and $\nu$ leads
to the two following forms, \b \label{Spa36} {a(\eta)}\approx
|\eta|^{\frac{1}{2}-{\nu}} \e or \b \label{Spa361}
{a(\eta)}\approx |\eta|^{\frac{1}{2}+{\nu}}. \e If we consider
$\nu=3/2$, we obtain for (\ref{Spa36}), ${a(\eta)}\approx
|\eta|^{-1}$ that indicates exponentially inflation in dS phase.
Equivalently, from (\ref{Spa36})and $dt=a(\eta)d\eta$, we can
drive scale factor in terms of cosmic time $t$ and $\nu$, \b
\label{Spa37} {a(t)}\approx
{t^\frac{\frac{1}{2}-\nu}{\frac{3}{2}-\nu}}. \e On the other hand,from Friedmann equation for hot standard model the following
expansion rate in terms of $t$ and equation of state $\omega$ is obtained \cite{inf2}, \b \label{Spa38} {a(t)}\approx
{t^\frac{2}{3(1+\omega)}}, \e with equality of two relations (\ref{Spa37}) and (\ref{Spa38}), we obtain the equation of state
in terms of index $\nu$ as follows: \b \label{Spa39}\omega=\frac{1}{3}\big({\frac{2\nu+3}{1-2\nu}}\big). \e
It means that the equation of state of the cosmic fluid, may be dependent
to the dynamics of background space-time.

\subsection{Initial Inflationary Modes in Terms of Index $\nu$}

The general form of mode equation (\ref{muk7}) for curved space-time is,
\begin{equation} \label{Muk25}  u''_{k}+(k^{2}-\frac{4\nu^2-{1}}{4\eta^2})u_{k}=0. \end{equation}
Consequently, according to the general equation of motion (\ref{Muk25}) and the asymptotic expansion (\ref{asm23}),
the general form of mode functions for any curved space-time becomes,
$$ u^{gen}_{k}(\eta,\nu)=A_{k}\frac{e^{-{i}k\eta}}{\sqrt{2k}}\big(1-{i}\frac{4\nu^2-1}{8k\eta}-\frac{(4\nu^2-1)(4\nu^2-9)}{2!(8k\eta)^2}-...\big)$$
 \begin{equation} \label{gen27}
 + B_{k}\frac{e^{{i}k\eta}}{\sqrt{2k}}\big(1+{i}\frac{4\nu^2-1}{8k\eta}-\frac{(4\nu^2-1)(4\nu^2-9)}{2!(8k\eta)^2}+...\big).
 \end{equation}
Note that, we consider $|\eta|=-\eta$ for very early time. Also, the general mode (\ref{gen27}) is a
function of both variables $\eta$ and $\nu$. The positive frequency solutions of the mode equation (\ref{Muk25}) becomes,
\begin{equation} \label{mod28}  u^{\nu}_{k}=\frac{e^{-{i}k\eta}}{\sqrt{2k}}\big(1-{i}\frac{4\nu^2-1}{8k\eta}-\frac{(4\nu^2-1)
(4\nu^2-9)}{2!(8k\eta)^2}-...\big)
. \end{equation}
We consider this solution as the \emph{non-dS(ND) modes} for general curved space-time in terms of index $\nu$.
Compared with usual BD mode
 \begin{equation} \label{mod282}  u^{BD}_{k}=\frac{e^{-{i}k\eta}}{\sqrt{2k}}\big(1-\frac{i}{k\eta}\big),
 \end{equation}
these general ND modes $u^{\nu}_{k}$, could be more complete
solution of the general wave equations (\ref{Muk25}) for the
general curved space-time (i.e. both of dS and non-dS space-time),
whereas BD mode is a specific solution for pure dS space-time.
Also, the BD mode is linear order of $\frac{1}{k\eta}$, but for
$\nu\neq\frac{3}{2}$, we must
have ND modes with terms that are non-linear order of $\frac{1}{k\eta}$.\\

\subsection{Power Spectrum in Terms of Index $\nu$}

To calculate the power spectrum, we need to compute the following quantity \cite{per5, non29},
\begin{equation}
 \label{pow16} \langle\hat{u}_{k}(\eta)\hat{u}_{k'}(\eta)\rangle=\langle0|\hat{u}_{k}(\eta)\hat{u}_{k'}(\eta)|0\rangle=\frac{1}{2}
 (2\pi)^{3})\delta^{3}(k+k')|u_{k}(\eta)|^2.
\end{equation}
In addition, we should introduce some standard quantities in terms of curvature perturbation ${\cal R}_{k}(\eta)$,
\begin{equation}
 \label{pow17} \langle\hat{{\cal R}}_{k}(\eta)\hat{{\cal R}}_{k'}(\eta)\rangle=(2\pi)^{3}\delta^{3}(k+k')P_{{\cal R}},
\end{equation}
\begin{equation}
 \label{pow18} \Delta_{{\cal R}}^{2}=\frac{k^3}{2\pi^{2}}P_{{\cal R}},
\end{equation}
where
\begin{equation}
 \label{pow19} {\cal R}_{k}(\eta)=\frac{u_{k}(\eta)}{z}=\frac{u_{k}(\eta)}{a}(\frac{H}{{\dot{\bar{\phi}}}}).
\end{equation}
$P_{{\cal R}}$ is the power spectrum and $\Delta_{{\cal R}}^{2}$ is the dimensionless power spectrum \cite{non29}.\\
By using equations (\ref{pow17} -\ref{pow19}), for ND modes (\ref{mod28}) up to second order of $1/k\eta$, the modified power spectrum obtain in the following form in terms of index $\nu$,
  \begin{equation} \label{del36}
 \Delta_{{\cal R}}^{2}=\frac{H^2}{(2\pi)^{2}}(\frac{H^2}{\dot{\bar{\phi}}^2})\big[\frac{2\nu+1}{2(2\nu-1)}+(2\nu+1)^{2}
  \frac{(4\nu^{2}-9)^{2}}{64k^{2}\eta^{2}}+...\big].
\end{equation}
Note that, we use conformal time $\eta$ in terms of index $\nu$ as\cite{asl55},
\begin{equation}
 \label{eta33} \eta=\frac{-1}{aH}\left(\nu-\frac{1}{2}\right).
\end{equation}
As another result of the general ND modes, we can study the spectra of created gravitational particles during the inflation in terms of index $\nu$ \cite{man24}.
\section{Discussion on Special Values of Index $\nu$ }

In this section, for a closer look at issues raised, we are going to study some special values of index $\nu$, such as, $\nu=\frac{\pm1}{2}, \frac{\pm3}{2}$, $\nu\simeq\frac{3}{2}\pm\epsilon$ with $\epsilon\ll1$, $\nu=\frac{5}{2}$ and $\nu=\frac{7}{2}$. It is worth to mention that, the ND modes (\ref{mod28}), regarding the Half-integer values of $\nu$ lead to the exact solutions and with respect to the another values, lead to the approximate ones.\\
 \begin{itemize}
  \item For $\nu=\frac{\pm1}{2}$, considering(\ref{zed24}), we have $ \frac{z''}{z}=0$. So,
  the positive mode functions for $\nu=\frac{\pm1}{2}$ lead to exact Minkowski mode as,
 \begin{equation}
 \label{Bun29} u_{k}^{\pm1/2}=\frac{e^{-ik\eta}}{\sqrt{2k}}.
\end{equation}
  \item In the case of $\nu=\frac{\pm3}{2}$, considering(\ref{zed24}), we have, $ \frac{z''}{z}=\frac{2}{\eta^2}$, and the general form of the positive mode functions lead to the exact BD mode:

  \begin{equation}
 \label{Bun293} u_{k}^{\pm3/2}=\frac{1}{\sqrt{2k}}(1-\frac{i}{k\eta})e^{-ik\eta}.
\end{equation}
For the special case $\nu=\frac{+3}{2}$, we have exponentially inflation,
$a(t)=e^{Ht}$ or $ a(\eta)=-\frac{1}{{H}\eta}$ with $H=constant$ for early universe.
  \item Also, considering $\nu=\frac{5}{2}$ and $\nu=\frac{7}{2}$, equation (\ref{zed24}) leads to,
  $ \frac{z''}{z}=\frac{6}{\eta^2}$ and $ \frac{z''}{z}=\frac{12}{\eta^2}$, respectively. So, we have exact solutions as,
  \begin{equation}
 \label{non mode} u_{k}^{5/2}=\frac{1}{\sqrt{2k}}(1-\frac{3i}{k\eta}-\frac{3}{k^2\eta^2})e^{-ik\eta},
\end{equation}
and
 \begin{equation}
 \label{non mode} u_{k}^{7/2}=\frac{1}{\sqrt{2k}}(1-\frac{6i}{k\eta}-\frac{15}{k^2\eta^2}+\frac{15i}{k^3\eta^3})e^{-ik\eta}.
\end{equation}
Note that these two last modes are the exact solutions of Mukhanov equation (\ref{Muk25}) and they are non-linear solutions
of order $\frac{1}{k^2\eta^2}$ and $\frac{1}{k^3\eta^3}$.

\item Since, inflation started in approximate dS space-time with varying $H$, basically in this Dynamical background of early universe,
finding a proper mode is difficult. So, we offer the general non-dS modes (\ref{mod28}), as the fundamental modes during
inflation that asymptotically approach to flat background in $\eta\rightarrow{-\infty}$. For these fundamental modes, we showed that the best values of index $\nu$ which are confirmed with the latest observational data, is $\nu=\frac{3}{2}+\epsilon$, where $0.01\lesssim \epsilon \lesssim 0.03$ \cite{man1}. For this range of index $\nu$, we have slow-roll power law inflation \cite{pli17}.
    \end{itemize}
Moreover, for the some special values of index $\nu$, we have;
$$\nu=\frac{1}{2}\Rightarrow\omega\rightarrow{\infty},$$
$$ \nu=\frac{-1}{2}\Rightarrow\omega=\frac{1}{3}(Radiation),$$
$$\nu=\frac{3}{2}\Rightarrow\omega=-1(Vacuum),$$ $$\nu=\frac{-3}{2}\Rightarrow\omega=0 (Presureless-Matter)$$.

By considering \emph{Slow-roll inflation}, i.e. $\nu=\frac{3}{2}+\epsilon$, we obtain $\omega=-0.99$ for $\epsilon=0.01\ll{1}$. Athwart,
if we consider $\nu=\frac{3}{2}-\epsilon$, we obtain $\omega=-1.1$, i.e. \emph{Phantom crossing limit}. Also for
\emph{dark energy fluid}, i.e.${-1}<\omega<\frac{-1}{3}$ \cite{ska1}, we have ${\frac{+3}{2}}<\nu<+\infty$.\\

At the end, we can obtain the form of the power spectrum for the early cosmological epochs as follows,\\
a) For the \emph{Flat space-time} ($\nu=\frac{1}{2}$), we obtain,
\begin{equation}
 \Delta_{{\cal R}}^{2}=0.
\end{equation}
b) For the \emph{de Sitter space-time} ($\nu=\frac{3}{2}$), we obtain conventional scale-invariant result as,
\begin{equation}
 \Delta_{{\cal R}}^{2}=\frac{H^2}{(2\pi)^{2}}(\frac{H^2}{\dot{\bar{\phi}}^2}).
\end{equation}
c) Finally for the \emph{Quasi-de Sitter space-time}
($\nu=\frac{3}{2}+\epsilon$), we obtain scale-dependent power
spectrum (\ref{del36}), that have been obtained theoretically by other researchers \cite{tra48,tra49, tra50, tra511, tra512} and have been confirmed by recent observations \cite{obs71, obs8}.\\

\noindent{\bf{Acknowlegements}}: We would like to thank M. V.
Takook, M. R. Tanhayi and H. Pejhan for useful and serious
discussions. This work has been supported by the Islamic Azad
University, Ayatollah Amoli Science and Research Branch, Amol,
Mazandaran, Iran.

\end{document}